\documentclass[11pt,a4paper]{article}
\pdfoutput=1
\usepackage{jheppub}
\usepackage{times}
\usepackage{graphicx}

\renewcommand{\Im}{{\rm Im}\,}
\renewcommand{\Re}{{\rm Re}\,}
\newcommand{\re}[1]{\Re\left(#1\right)}
\newcommand{\im}[1]{\Im\left(#1\right)}
\newcommand{\mathds}[1]{}

\renewcommand{\sp}[2]{\left\langle  \vphantom{#1}\vphantom{#2} #1\right.\left.\vphantom{#1}\vphantom{#2} \right| \left. #2  \vphantom{#1}\vphantom{#2} \right\rangle}

\newcommand{\V}{{V}}
\newcommand{\K}{\mathcal{K}}
\newcommand{\R}{\mbox{$\mathcal{R}$}}
\newcommand{\I}{\mbox{$\mathcal{I}$}}

\newcommand{\G}{\mathcal{G}}

\newcommand{\Pp}{\text{P}_{\scriptscriptstyle{+}}}
\newcommand{\Pm}{\text{P}_{\scriptscriptstyle{-}}}
\newcommand{\Ppm}{\text{P}_\pm}
\newcommand{\Pmp}{\text{P}_\mp}

\newcommand{\vecr}{\mathbf{x}}

\newcommand{\ssigmaad}{{\cal S}^\dagger}
\newcommand{\ssigma}{{\cal S}}
\newcommand{\atot}{\mbox{$A_\infty$}}
\newcommand{\ah}{\mbox{$S_h$}}

\begin{document}

\begin{flushright}
{\small
FISPAC-TH/13-314\\
UQBAR-TH/78-156\\
\today\\
\normalsize}
\end{flushright}

\begin{center}

\vspace{1.5cm}

{\LARGE \bf  N=2 SUGRA BPS Multi-center solutions,
  quadratic\\[0.2cm] 
prepotentials and Freudenthal transformations  }


\vspace{0.7cm}

\begin{center}

{\sl\large J.J.~Fern\'andez-Melgarejo$^{\dagger}$}
\footnote{E-mail: {\tt jj.fernandezmelgarejo@um.es}},
{\sl\large and E.~Torrente-Lujan$^{\dagger}$}
\footnote{E-mail: {\tt etl@um.es,torrente@cern.ch}}

\vspace{0.2cm}

${}^{\dagger}${\it Departamento de F\'{\i}sica, U. de Murcia, Campus de Espinardo, 30100 Murcia, Spain}\\ 

\vspace{.2cm}

\end{center}
\vspace{0.3cm}
{\bf Abstract}

\begin{quotation}
{
We present a detailed 
 description of $N=2$ stationary 
BPS multicenter black hole solutions for quadratic prepotentials
with an arbitrary number of centers and scalar fields
making a systematic  use of 
the algebraic properties of the matrix of second derivatives
of the prepotential,  $\mathcal{S}$, which in this case is
a scalar-independent matrix.
In particular we obtain bounds on the physical parameters 
of the multicenter solution such as horizon areas and ADM 
mass. 
We discuss the possibility and convenience of setting up a 
basis of the symplectic vector space built from charge 
eigenvectors of the $\ssigma$, the set of vectors 
$(\Ppm q_a)$ with $\Ppm$ $\ssigma$-eigenspace projectors.

The anti-involution matrix $\mathcal{S}$ can be understood 
as a  Freudenthal duality $\tilde{x}=\ssigma x$.
We show that this duality can be generalized to
``Freudenthal transformations'' 
$$x\to \lambda\exp(\theta \ssigma) x= a x+b\tilde{x}$$
under which the horizon area, ADM mass and intercenter 
distances scale up leaving constant the scalars at the fixed points.
In the special case $\lambda=1$, ``$\ssigma$-rotations'', 
the transformations leave invariant the solution.
The standard Freudenthal duality can be written as
 $\tilde x= \exp\left(\frac{\pi}{2} \ssigma\right) x .$
We argue that these generalized transformations  leave
invariant not only the quadratic prepotential theories
 but also
 the general stringy extremal quartic form $\Delta_4$,
$\Delta_4(x)= \Delta_4(\cos\theta x+\sin\theta\tilde{x})$
and therefore its entropy at lowest order.

\vspace{0.7cm}
}
\end{quotation}
\end{center}

\newpage
\section{Introduction}
\label{ch:bh}


We present a systematic study of extremal, stationary, multi-center black-hole-type solutions in $N=2$ $D=4$ ungauged Einstein-Maxwell  supergravity theories minimally coupled to an arbitrary number $n_v$ of vector multiplets, i.e. quadratic prepotentials. 



The action of these $4D$ $N=2$ supergravities can be written, 
in the framework of special geometry, in terms of a 
holomorphic section $\Omega$ of the scalar manifold. The 
corresponding 
field equations and Bianchi identities remain invariant under 
the group of symplectic transformations 
$Sp(2n_v+2,\mathbb{R})$. This group acts linearly on the 
section $\Omega$, which 
transforms as a symplectic vector when it is parametrized as $\Omega=(X^I,F_I)$, for $I=0,\ldots,n_v$. 


%
The embedding of the duality group of the moduli space into the symplectic group $Sp(2n_v+2,\mathbb{R})$ establishes, in general, a relation between the upper and lower components of $\Omega$, $X^I$ and  $F_I=F_I(X^J)$ respectively. In some cases, $F_I$ is the derivative of a single function, the prepotential $F=F(X^J)$. The choice of a particular embedding determines the full Lagrangian of the theory and whether a prepotential exists \cite{Sabra:1996xg,Sabra:1996kw}.

In this work, we restrict ourselves to 
 general quadratic prepotentials. 
These theories 
\footnote{See \protect\cite{Cremmer:1984hc} for a classification of $N=2$ SUGRA special K\"ahler manifolds.}
include the simplest examples of special K\"ahler homogeneous manifolds, the axion-dilaton model or  the
\begin{align}
\mathbb{C}P^n\equiv \frac{SU(1,n)}{U(1)\times SU(n)}
\end{align}
case.

%
These models correspond to Einstein-Maxwell $N=2$ supergravities  minimally coupled to $n_v$ vector multiplets. They lead to phenomenologically interesting $N = 1$ minimally coupled supergravities \cite{Ferrara:2012qp}.
Theories derived from particular examples of these quadratic prepotentials have been studied in detail.
\footnote{The case $n_v=1$ corresponds to the $SU(1,1)/U(1)$ axion-dilaton black hole \protect\cite{Sabra:1996bk,Behrndt:1997fq,Ferrara:2010cw} with
prepotential $ F=-i X^0X^1$.
}

%
%

Black hole solutions in $N=2$ $D=4$ supergravity have been
 extensively studied for a long term by now. 
See, for example, refs. \cite{Ferrara:1995ih,Ferrara:1996um,Gibbons:1996af,Shmakova:1996nz,Ferrara:1997tw}, 
\cite{Behrndt:1997ny,Sabra:1997dh,Sabra:1997kq,Bellucci:2007ds,Ceresole:2010hq,Andrianopoli:1996vr,Galli:2011fq,Meessen:2006tu}. 
Multicenter black holes have been treated in refs.
\cite{Liu:2000ah,Bellorin:2006xr,Bobev:2009zz,Bena:2009en,Bena:2008wt,Anninos:2011vn,Manschot:2011xc}.
In this work we show how it is possible a detailed study 
of stationary multicenter black-hole type solutions 
with any number of scalar fields and centers, of the properties of the bosonic field solutions and their 
global and local properties
making a systematic and intensive use of 
the algebraic properties of the matrix of second derivatives
of the prepotential, the matrix $\mathcal{S}$
and of the matrix $\ssigmaad$, its adjoint with respect the 
symplectic product.
This matrix 
is an isometry of the symplectic bilinear form, 
it connects 
 the real and imaginary parts of symplectic sections of the 
theory. In this case it is a real 
scalar-independent  $Sp(2 n_v+2,\mathbb{R})$ 
matrix. 
Among other results, 
 we obtain bounds  the physical parameter 
of the multicenter solution such as horizon areas and ADM 
mass valid for any quadratic prepotentials. 

The compatibility of the matrix $\ssigma$
 with respect to the symplectic product makes possible the definition of an associated inner product for which these matrices are unitary. 
We discuss the possibility and convenience of setting up a 
basis of the 
$(2 n_v+2)$-dimensional symplectic 
 vector space built from charge 
eigenvectors of the matrix $\ssigma$.  This
 set of vectors are of the form
$(\Ppm q_a)$, 
or, alternatively, $(q_a,\ssigma q_a)$,   
with $\Ppm$  projectors over the eigenspaces of $\ssigma$
and $q_a$ the center vector charges.

The anti-involution matrix $\mathcal{S}$ can be understood 
as a  Freudenthal duality $\tilde{x}=\ssigma x$
\cite{Borsten:2009zy,Ferrara:2011gv}. 
We will show here that this duality can be generalized to an Abelian group of transformations
$$x\to \lambda\exp(\theta \ssigma) x= a x+b\tilde{x}.$$
Under this set of transformations applied to 
the charge vectors and $\I_\infty=\I(r\to\infty)$, 
 the horizon area, ADM mass and intercenter 
distances scale up, respectively, as 
$$\ah\to\lambda^2\ah,\quad M_{ADM}\to\lambda M_{ADM},\quad r_{ab}\to\lambda r_{ab}$$
leaving invariant the values of the scalars at the fixed points and at infinity.
In the special case $\lambda=1$, ``$\ssigma$-rotations'', 
the transformations leave invariant the solution. 
The standard Freudenthal duality can be written as
the particular rotation
 $$\tilde x= \exp\left( \frac{\pi}{2} \ssigma\right) x\,  .$$
We argue at the final section of this work 
that these generalized Freudenthal transformations leave 
invariant not only the entropy and other 
macroscopical quantities of quadratic prepotential theories
but also
 $\Delta_4$,
$\Delta_4(x)= \Delta_4(\cos\theta x+\sin\theta\tilde{x})$,
the quartic invariant 
\cite{Borsten:2009zy}
appearing in the description 
of more general theories, $4d$ SUGRAs that arise from 
String and M-theory and therefore the lowest order 
entropy of these theories.


In  Section \ref{ssection3}, we present some well-known basic aspects of $N=2$ $D=4$ supergravity theories and their formulation in terms
of special and symplectic geometry.
In Section \ref{ssection4},  we first introduce the matrices
$\ssigma_{N,F}$, stressing some of their known properties and deriving new ones. We also construct projective operators (as well as their corresponding symplectic adjoints) based on these matrices.
After the consideration of the attractor mechanism in terms of these projectors, we enter in a full explicit description of multicenter black hole solutions, their horizons and their asymptotic properties. This is done in Sections \ref{ssection5} and \ref{ssection6}.
We finally present Section \ref{ssection7}, which contains a summary and discussion of our work, as well as an outlook on further proposals.


\section{$N=2$ $D=4$ SUGRA and Special K\"ahler geometry}
\label{ssection3}

The field content of the
$N=2$ supergravity theory coupled to $n_v$ vector multiplets
consists of
\begin{align}
\left\{
	e_\mu{}^a \, ,
	A_\mu{}^I \, ,
	z^\alpha \, ,
	\psi_\mu{}^r \, ,
	\lambda_r{}^\alpha
	\right\}
\, ,
\end{align}
with $\alpha=1,\ldots, n_v$, and $I=0,\ldots,n_v$.
The theory
also contains some hypermultiplets, which can be safely  taken
as constant or neglected (further
details can be found in \cite{Bellorin:2006xr}, whose notation and concepts we generally adopt).
The   bosonic $N=2$ action  can be written as
\begin{align}
S&=\int_{M(4d)} R\star 1
+\G_{\alpha\bar\beta} d z^\alpha\wedge \star d\bar z^{\bar\beta}
+  F^I\wedge  G_I
\, .
\end{align}
The fields $F^I,G_I$ are not independent.
Whilst $F^I$  is given by $F^I=d A^I$, $G_I$ is a set of combinations
of the $F^I$ and their Hodge duals,
\begin{align}
G_I=a_{IJ} F^I+b_{IJ}\star F^I
\, ,
\end{align}
with scalar-dependent coefficients $a_{IJ}$ and $b_{IJ}$.

Abelian charges with
respect the $U(1)^{n_v+1}$ local symmetry of the
theory are defined by means of the integrals of the gauge field strengths.
The total charges of the geometry  are
\begin{align}
q
&\equiv (p^I,q_I)
\equiv  \frac{1}{2\pi^2}\int_{S_\infty} ( F^I, G_I)
\, .
\label{eq22}
\end{align}
Similar charges can be defined for specific finite regions.

The theory is defined, in the special geometry formalism,
by the introduction of some projective scalar coordinates $X^I$,
as for example,  `special' projective  coordinates
$ z^\alpha\equiv X^\alpha/X^0$.  By introducing a covariantly holomorphic section of a symplectic bundle, $V$, we are able to arrange $2n_v$ quantities that transform as a vector under symplectic transformations at any point of the manifold. $V$ has the following structure
$V=V(z,{\bar z}) \equiv (V^I,V_I)$
and satisfies the following identities:
\begin{align}
\sp{V}{\bar V}
&\equiv V^t {\cal \omega}\bar V
\equiv {\bar V}^I V_I-V^I {\bar V}_I
=-i
\, ,
\end{align}
where $\omega$ is the symplectic form.\footnote{We
choose a basis such that $\omega = \left( \begin{array}{cc}  0 & -{1}_{n_v} \\ {1}_{n_v} & 0 \end{array} \right)$.}

The scalar kinetic term in the action can be written in terms 
of $\V$ as 
$L_{s,kin}\sim  i \sp{D^\mu\bar{\V}}{D_\mu\V}$ 
and the scalar  metric is given by
\begin{align}
\G_{\alpha\bar\beta}= \partial_\alpha \partial_{\bar\beta}\K
\, ,
\end{align}
where the K\"ahler potential $\K$ is defined by the relations
$V=\exp({\cal K}/2) \Omega$ being $\Omega \equiv (X^I, F_I)$
a holomorphic section
and
\begin{align}
e^{-{\cal K}}&=i \left ({\bar X}^I F_I- X^I \bar{F}_I\right ) =
i \sp{\Omega}{\bar \Omega}.
\label{eq697aa}
\end{align}
In $N=2$ theories, the central charge $Z$  can be expressed as
a  linear function on the charge space:
\begin{align}
\label{eq26}
Z(z^\alpha,q)\equiv\sp{V}{q}=e^{\K/2} \left(p^I F_I-q_I X^I\right)
 \, .
\end{align}

The embedding of the isometry group of the scalar manifold
metric $\G_{\alpha\bar\beta}$, into the symplectic group
fixes, through the K\"ahler potential $\K$, a functional 
relation between the lower and upper parts of $V$ 
and $\Omega$ \cite{hepth9611210,hepth9608075},
\begin{align}
F_I&=F_I(X^I)\, ,\label{eq74a}\\
V_I&=V_I(V^I)\, .
\label{eq74b}
\end{align}

There always exists a symplectic frame under which the theory 
can be described in terms of a single holomorphic function,
the \emph{prepotential} $F(X)$. It is a second degree 
homogeneous function on the projective scalar coordinates 
$X^I$, such that $F_I(X)=\partial_I F (X) $. For simplicity, 
we will assume the existence of such prepotential along this 
study although the results will not depend on such existence. 
Using the notation $F_{IJ}= \partial_I\partial_J F$,
the lower and upper components of $\Omega$  are related by
\begin{align}
F_I &= F_{IJ} X^J\, .
\end{align}

The lower and upper components of $V$  are related by a field 
dependent matrix $N_{IJ}$, which is determined by the special 
geometry relations
\cite{Ferrara:1996um}
\begin{eqnarray}
V_I &=& N_{IJ} V^J,\quad \label{eq24}\\
D_{\bar\imath} {\bar V}_I &=& N_{IJ} D_{\bar \imath} {\bar V}^J.
\label{eq24b}
\end{eqnarray}

The matrix $N$, which also fixes the vector couplings $(a_{IJ},b_{IJ})$ in the action, can be related  
to $F_{IJ}$ \cite{Ceresole:1995ca} by
\begin{align}
N_{IJ}={\bar F}_{IJ}+ T_I T_J\, ,
\end{align}
where the quantities $T_I$ are proportional to the projector 
of the graviphoton, whose flux defines the $N=2$ central 
charge \cite{Ceresole:1995ca}. For our purposes, it is 
convenient to  write the relation between  $N_{IJ}$ 
and $F_{IJ}$ as
\begin{align}
N_{IJ} &\equiv F_{IJ}+N_{IJ}^\perp
\nonumber\\
&=F_{IJ}- 2i \im{F_{IJ}}+2 i\frac{ \im{F_{IK}} L^K \im{F_{JQ}} L^Q}{L^P\im{F_{PQ}} L^Q}
\, ,
 \end{align}
where we have decomposed the matrix $N_{IJ}$
into   ``longitudinal''  (the $F_{IJ}$ themselves)
and  ``transversal'' parts  ($N_{IJ}^\perp$).
The perpendicular term (defined by the expression above)
  annihilates $L^I$, or any multiple of it,
\begin{eqnarray}
N_{IJ}^\perp (\alpha L^J) &=& 0
\, .
 \end{eqnarray}
From this, (\ref{eq24}) can be written as
\begin{align}
V_I & =N_{IJ} L^J = \left (F_{IJ}+ N_{IJ}^\perp \right) L^J
\nonumber\\
    &=         F_{IJ}  L^J\, .
\label{eq786}
 \end{align}
Thus, the upper and lower components of $V$ and $\Omega$
are connected by the same matrix $F_{IJ}$.

The existence of  functional dependencies among the upper 
and lower components of the vectors $V$ or  $\Omega$
 imply further relations between their respective   real 
and imaginary parts. They are related by symplectic
matrices $\ssigma(N),\ssigma(F)\in Sp(2 n_v+2,\mathbb{R}) $ 
which are respectively associated to the 
quantities $N_{IJ},F_{IJ}$ as follows: 
\begin{align}
\re{\Omega}&=\ssigma(F)\im{\Omega}
\, ,
\label{eq234a}
\\
\re{V}&=\ssigma(N)\im{V}=\ssigma(F)\im{V}
\, .
\label{eq234}
\end{align}
The last expression is obtained by means of the relation (\ref{eq786}). These same relations \eqref{eq234a}-\eqref{eq234}
are valid for any complex multiple of $\Omega$ or $V$.
It is straightforward to  show, for example, that for
 any $\lambda\in\mathbb{C}$, we have
\begin{align}
\re{ \lambda V}&=\ssigma(N)\im{ \lambda V}=
\ssigma(F)\im {\lambda V}
\, .
\label{eq122}
\end{align}


%
The matrix  $\ssigma (F)$ is,
by direct computation (Eq.(75) in \cite{Ceresole:1995ca}), of the form 

\begin{align}
\mathcal{S}(F) &=
\left( \begin{array}{cc}
1 & 0\\
\re {F_{IJ}}  & -\im{F_{IJ}}
\end{array} \right)
\left( \begin{array}{cc}
0 & 1 \\
\im {F_{IJ}} & \re {F_{IJ}}^{-1}
\end{array} \right)^{-1}
\, .
\label{eq680bb}
\end{align}
Similarly, the same result applies for $\mathcal{S}(N)$ with $F_{IJ}\to N_{IJ}$.\footnote{The matrix $\ssigma_N$ is related to $\mathcal{M}$, the matrix that appears in the black hole effective potential \protect\cite{Ceresole:1995ca}
$V_{BH}=-\tfrac{1}{2} q^t{\cal M}q $,  by
${\cal\omega}\ssigma(N)={\cal M}$.}


In $N=2$ theories, $\ssigma(N)$ always exhibits a moduli dependence \cite{Ferrara:2012qp}.
However, this is not the case for $\ssigma(F)$.
We will  focus in this work
on the particular case of theories with
quadratic prepotentials,
\begin{align}
 F(X) &= \frac{1}{2} F_{IJ} X^IX^J
 \, ,
\end{align}
where  $F_{IJ}$ is a complex, constant, symmetric matrix.
Then, the corresponding matrix $\ssigma(F)$
is a field-independent,  constant matrix.
We can assume that $\re{F_{IJ}}=0$  and $\im{F_{IJ}}$ is
 negative definite.
In what follows, we will use the notation $\ssigma\equiv \ssigma(F)$.
The condition $e^{-\K}>0$ and the expression 
\eqref{eq697aa}
implies a restriction on the prepotential. We will write
this restriction in a convenient form in 
Section \ref{ssection4} in terms of the positivity of 
a quadratic form.


\subsection{General supersymmetric stationary  solutions}

The most general stationary (time independent) 4-dimensional metric compatible with supersymmetry can be
written in the IWP form
\cite{Israel:1972vx,Perjes:1971gv,Sabra:1997yd},
\begin{align}
 ds^2&=e^{2 U}(dt+\omega)^2-e^{-2 U} d{\vecr}^2.
\label{eq211}
\end{align}

Supersymmetric $N=2$ supergravity solutions can be constructed
systematically following well established methods
\cite{Meessen:2006tu,Bellorin:2006xr}. 
In this section we will closely follow the 
notation of ref.\cite{Bellorin:2006xr}. 
The 1-form $\omega$ and the function $e^{-2U}$ are related in
these theories to the K\"ahler potential and
connection, $\K,Q$ \cite{Sabra:1997yd}.
We demand
 asymptotic flatness, $e^{-2 U}\to 1$
 together with $\omega\to 0$
for $|\vecr|\to\infty$.
BPS field equation solutions for the action above (for example, quantities that appear in the metric, as $e^{-2U}$ or $\omega$) can be written
in terms of the following real symplectic vectors $\R$ and $\I$
\begin{align}
\R&=\frac{1}{\sqrt 2}\re{\frac{V}{ X}} \, ,
\\
\I&=\frac{1}{\sqrt 2}\im{ \frac{V}{ X}}\, .
\end{align}
$X$ is an arbitrary complex function of space coordinates
 such that $1/X$ is harmonic.
The $2 n_v+2$ components of $\I$ and $\R$ are real harmonic functions
in $\mathbb{R}^3$.
There is an algebraic
 relation between $\R$ and $\I$ and
the solutions can be written in terms only of the  vector $\I$.
By making use of  \eqref{eq234a}-\eqref{eq122}, we can write the following
\emph{stabilization equation}
\begin{align}
\R &= \ssigma \I
\, .
\label{eq888}
\end{align}

In practice,
specific solutions are determined by giving a
particular, suitable, ansatz for the symplectic vector $\I$ as
a function of the spacetime coordinates.


Using these symplectic vectors we rewrite the
only independent metric component as
\begin{align}
e^{-2 U} &=e^{-\K}= \frac{1}{2 | X|^2}
\nonumber\\
               &=  \sp{\R}{\I}=\sp{\ssigma\I}{\I}
\, .
\label{eq215}
\end{align}
Similarly,  the time independent $3$-dimensional 1-form  $\omega=\omega_i dx^i $ satisfies the equation
\begin{align}
 d\omega=2\sp{\I}{\star_3 d\I}
 \, ,
 \label{eq687bb}
\end{align}
where $\star_3$ is the Hodge dual on flat $\mathbb{R}^3$, together with
 the integrability condition
\begin{align}
 \sp{\I}{\Delta\I}&=0
 \, .
\label{eq221}
\end{align}
The asymptotic flatness condition implies
\begin{align}
 \sp{\R_\infty}{\I_\infty}&=\sp{\ssigma\I_\infty}{\I_\infty}=1
 \, .
\label{eq689a}
\end{align}

The gauge field equations of motion and Bianchi identities
 can be directly solved  in terms of spatially
dependent harmonic functions \cite{Bellorin:2006xr}.
The modulus of the central charge function defined in
\eqref{eq26} can be written,  taking into account
\eqref{eq215}, as
\begin{eqnarray}
| Z(q)|^2 e^{- 2U} &=&| \sp{\R}{q}|^2+ | \sp{ \I}{q}|^2
\, .
\end{eqnarray}
At spatial infinity, by assuming the asymptotic flatness condition
\eqref{eq689a}, we arrive to
\begin{eqnarray}
| Z_\infty(q)|^2   &=&| \sp{\R_\infty}{q}|^2+ | \sp{ \I_\infty}{q}|^2
\, .
\label{eq218b}
\end{eqnarray}

The, assumed time independent, $n_v$ complex scalar fields
$z^\alpha$ solutions to the field equations, are given in
this formalism by
\begin{eqnarray}
 z^\alpha&=&\frac{\Omega^\alpha}{\Omega^0}=\frac{V^\alpha/X}{\V^0/X}=
\frac{\R^\alpha+i\I^\alpha}{\R^0+i\I^0}
\, .
\label{eq226}
\end{eqnarray}
This is, in general, a formal expression as the $\I$ or $\R$ quantities
may be scalar dependent.\footnote{Even for a scalar independent
ansatz $\I$, the matrix $\ssigma$ is, in general, scalar dependent.}

These scalar fields can, in principle, take any values
($z_\infty$) at infinity. These values will
appear as free parameters in the ansatz that we give for $\I$.
Nevertheless, according to the attractor mechanism,
the moduli adjust themselves  at some fixed points.


We are interested in this work in extremal, single- or
multi-center black hole-type solutions  determined by
an $\I$ ansatz with point-like singularities of the form
\begin{align}
 \I&=\I_{\infty}+\sum_{a}\frac{q_a}{|\vecr-\vecr_a|}
 \, ,
\label{eq216}
\end{align}
where $a=1,\ldots,n_a$ being the number $n_a$ arbitrary and
$q_a=(p_a{}^I, q_{aI})$ and $\I_\infty$ real, constant,
symplectic vectors.

For this kind of solutions, the quantities
$\I_\infty$ are related to the values at infinity of
the moduli while the ``charge'' vectors $q_a$ are
related to their values at the fixed points.
The fixed values of the scalars, $z(\vecr)\to z(\vecr_a)=z_{f}^a$, are the solutions
of the following
\emph{attractor equations}
\cite{Shmakova:1996nz,Ferrara:1995ih,Ferrara:1996um}:
\begin{align}
q^a=\re{2i\bar Z(z_{f}^a) V(z_f^a)}
\, .
\label{eq224}
\end{align}
The prepotential performs its influence throughout $V$ and $Z$ (\emph{cf.} \eqref{eq26}).
The scalar attractor values are independent of their
asymptotic values and only depend on the discrete charges
$ z^a_{f}=z^a_{f}(q_a)$.

Single center black hole solutions are known to exist for all
regions of the moduli scalars at infinity, under very mild
conditions on the charge vector. In the multicenter case,
for fixed charge vectors, not all the positions $\vecr_a$
 in the ansatz \eqref{eq216} are allowed.
The integrability condition (\ref{eq221}) imposes
necessary conditions on the relative positions and on the moduli
at spatial infinity (through $\I_\infty$)
for the existence of a solution.
In this framework, a particular black hole solution is
completely determined by a triplet of charge vectors,
distances and values of the moduli at
infinity $(q_a,\vecr_a,z^\alpha_\infty)$.

\section{The stabilization matrix, its adjoint and the attractor equations}
\label{ssection4}
\label{sec:stabilizationequations}
Let us consider now the attractor equations (\ref{eq224})
in more detail. We will use the properties of the
stabilization matrix $\ssigma$ to solve them in a purely
algebraic way to obtain some properties and give some explicit
expressions for the scalars at the fixed points.

For this purpose, we first establish some well-known
properties of  $\ssigma_N,\ssigma\equiv\ssigma_F$ and
define new matrices: some projector operators associated
to them and their respective symplectic adjoints.

It can be shown by explicit computation that the real
symplectic  matrices
$\ssigma_N,\ssigma\equiv\ssigma_F\in Sp(2n_v+2,\mathbb{R})$ defined by
\eqref{eq234a}-\eqref{eq234},
whose explicit expressions are \eqref{eq680bb},
  satisfy the relations (see also \cite{Ceresole:1995ca} )
\begin{eqnarray}
\ssigma_N^2=\ssigma_F^2&=&-{1}
\, .
\label{eq235}
\end{eqnarray}
From this, it is possible and convenient
 to define  the projector operators
\footnote{This is done for $\ssigma\equiv \ssigma_F$, but a similar procedure can be done for $\ssigma_N$.}
\begin{align}
 \Ppm = \frac{1\pm i\ssigma}{2}
 \, .
\end{align}
They satisfy the following straightforward properties
\begin{align}
 \Ppm^2       &=   \Ppm\, , \nonumber\\
 \ssigma \Ppm &= \mp i  \Ppm\, , \label{eq311}\\
 (\Ppm )^*    & =\Pmp\, . \nonumber
\end{align}
For $X$, $Y$ arbitrary real vectors, we have
\begin{align}
 \Ppm X    & =\Ppm Y \quad \Rightarrow\, \quad X=Y\, .  
\label{eq312cc}
\end{align}

According to \eqref{eq311}, $\Ppm$ are the
projectors on the eigenspaces (of equal dimension) 
of the matrix $\ssigma$.
The symplectic space $W$ can be decomposed into
eigenspaces of the matrix $\ssigma$:
\begin{align}
 W &= W^+\oplus W^-\, ,
\end{align}
where $W^{\pm} \equiv \Ppm  W$.
For an arbitrary function of $\ssigma$, $f(\ssigma)$, 
necessarily a linear function of it, 
$$f(\ssigma)\equiv a+b \ssigma \equiv \lambda \exp (\theta \ssigma),$$ we have 
\begin{eqnarray}
f(\ssigma)\Ppm&=&f(\mp i )\Ppm.
\label{eq7778}
\end{eqnarray}
Complex conjugation interchanges $W^+$ and $W^-$ subspaces, both subspaces are isomorphic to each other and of dimension
$n_v+1$.

We can rewrite a stabilization relation for the projectors $\Ppm$ analogous to \eqref{eq122}.
For arbitrary $\lambda\in \mathbb{C}$ and $V\in W$, for which there is a relation between
its real and imaginary parts of the form
$\re{V}=\ssigma\im{V}$, we have
 \begin{align}
\lambda V&=  \re{\lambda V}+i \im{\lambda V} =  
 2 i \Pm\im{\lambda V}\, .
\label{eq41}
\end{align}
Thus, the full vector $V$ can be reconstructed applying one of the projectors either from its real or imaginary
part.
We see that such vectors are fully contained in the
subspace $W^-$ or, equivalently, they are
eigenvectors of $\ssigma$
 \begin{align}
\ssigma V
&= 
2 i\ssigma   \Pm\im { V}
=  2  \Pm \im { V}
=    i V
\, .
\label{eq41cc}
\end{align}


We find it convenient to define the adjoint operator $\ssigmaad$ of the
matrix $\ssigma$, with respect to the symplectic
bilinear product so that, for any  vectors $A,B\in W$, we have
\begin{equation}
 \sp{\ssigma A}{B}=\sp{A}{\ssigmaad B}
 \, .
\end{equation}
A straightforward computation shows that  $\ssigmaad$ is given by
\begin{equation}
 \ssigmaad=-\omega \ssigma^t\omega
 \, .
\end{equation}
Under the assumption of a symmetric $F_{IJ}$ matrix, it is given by
\begin{align}
 \ssigmaad &= -\ssigma
 \, .
\label{eq1000}
\end{align}
In summary, the matrix $\ssigma$ is skew-adjoint with respect
to $\omega$ and its square is $\ssigma^2=-{1}$.
It fulfills an ``unitarity'' condition 
$\ssigmaad \ssigma={1}$.

$\ssigma$ defines an
(almost) complex structure on the symplectic space.
This complex structure preserves the symplectic bilinear form,
the matrix $\ssigma$ is an isometry of the symplectic space,
\begin{eqnarray}
 \sp{\ssigma X}{\ssigma Y}&=&\sp{X}{Y}
 \, .
\label{eq316}
\end{eqnarray}
From \eqref{eq1000}, we see that $\ssigma$ is an
element of the symplectic Lie algebra $\mathfrak{sp}(2 n_v+2)$.

Moreover, the bilinear form defined by
\begin{align}
g(X,Y)\equiv \sp{\ssigma X}{Y}
\end{align}
is symmetric.\footnote{The quadratic form $g(X,X)$ is also 
known as the ``$\I_2(X)$'' in the literature.
The corresponding quartic invariant in this case 
can be written as $\I_4(X,...)=\sp{X}{X}^2$.}
This can be easily seen as (using $\ssigmaad=-\ssigma$):
\begin{align}
 g(X,Y)&= \sp{\ssigma X}{Y}=\sp{X}{\ssigmaad Y}=\sp{\ssigma Y}{X}
 \nonumber\\
 &= g(Y,X)
 \, .
\label{eq1123b}
\end{align}


We will apply these properties to the study of the attractor
equations. In general, the matrices
$\ssigma_N,\ssigma_F$ are scalar dependent.
Only one of them, $\ssigma_F$, is constant,
in the case of quadratic prepotentials.
Let us write $\ssigma_N^f\equiv\ssigma_N(z_f)$
$\ssigma_F^f\equiv\ssigma_F(z_f)$ for the matrices
evaluated at (anyone of) the fixed points. Let us use the sub/superindex $f$ to denote any quantity at the fixed points. For instance, $Z^f\equiv Z(z^\alpha_f)$ or $V^f\equiv V(z^\alpha_f)$.
If we multiply both sides of
\eqref{eq224} by $\ssigma_N^f\equiv\ssigma_N(z_f)$,
 we arrive to
\begin{align}
\ssigma_N^{f} q^a&=
\ssigma_N^{f}\re{2i\bar Z^f V^f}=\ssigma_F^{f}\re{2i\bar Z^f V^f}
=\ssigma q^a
\, ,
\label{eq323}
\end{align}
where we have used \eqref{eq234} and (\ref{eq122}).\footnote{Following \cite{Ceresole:1995ca}, 
we note that
$V_{BH}=| Z_i|^2+| Z|^2=-\tfrac{1}{2} q^t \ssigma(N)\omega q$
 and
$
| Z_i|^2-| Z|^2=\frac{1}{2} q^t \ssigma(F)\omega q
$.
At the fixed points, we have $Z_i=0$, so that
$| Z|^2=-\frac{1}{2} q^t \ssigma_N \omega q=-\frac{1}{2} q^t\ssigma_F\omega  q $.
This last equation is satisfied by a solution of \eqref{eq224}.
Moreover the symmetric matrix $\omega \ssigma$ is indefinite, it has positive and negative eigenvalues.}

The  attractor equations can be written yet in another
alternative way. By using
(\ref{eq41}) and (\ref{eq224}), we can write
\begin{align}
i\bar Z^{f} V^{f}
&= 2 \Pm i\bar Z^{f} V^{f}
\nonumber\\
&= \Pm q,
\label{eq48}
\end{align}
or  its conjugate equation
\footnote{These equations are well  known in the literature, 
see for example section (5 ), Eq.(319), in \protect\cite{Youm:1997hw} and references therein. }
\begin{eqnarray}
-i Z^{f} \bar V^{f}&=& \Pp q
\, .
\label{eq311zz}
\end{eqnarray}
That is, the attractor equations simply
equalize (a multiple of) the vector $V$ (which, as we have seen above, lies
in the subspace $W^-$) with the part of the charge
vector which lies in $W^-$.

From \eqref{eq48}-\eqref{eq311zz}, by taking symplectic
products, we obtain
\begin{eqnarray}
|Z^{f}|^2 \sp{ V_{f}}{\bar V_{f}}
      &=&\sp{\Pm q}{\Pp q}=\sp{q}{\Pp q}\nonumber\\
      &=& -\frac{i}{2} \sp{\ssigma q}{ q}
      \, .
\end{eqnarray}
If we plug the constraint $\sp{V}{\bar V}=-i$,
we arrive in a straightforward and purely algebraic way
 to the  well known formula
\begin{eqnarray}
 |Z^{f}|^2&=& \frac{1}{2}\sp{\ssigma q}{q}
 \, ,
\label{eq412}
\end{eqnarray}
which relates the absolute value
of the central charge at any fixed point to a quadratic
expression on the charge.
It is obvious from \eqref{eq412} that the
 positivity of the quadratic form $g(q,q)=\sp{\ssigma q}{q}$
(at least locally at  all  the fixed points)
is  a necessary consistency condition for the
existence of solutions to the attractor mechanism.

Moreover the  
consistency condition $e^{-\K}>0$ 
can be written as (see \eqref{eq697aa}) 
\begin{eqnarray}
e^{-\K} &=& i \sp{\Omega}{\bar\Omega}=2 
\sp{\re{\Omega}}{\im{\Omega}}\nonumber\\
 &=&2 \sp{\ssigma\im{\Omega}}{\im{\Omega}}>0.
\label{eq6150b}
\end{eqnarray}
This condition is not  automatically 
satisfied as the symmetric quadratic form $g$ is indefinite.
\footnote{ The matrix $\omega\ssigma$ it has an even number 
of negative eigenvalues as
 $\det \omega\ssigma=(-1)^{2 n_v+2}=1$. The signature 
of $g$ is $(2 n_v,2)$.
}
In addition to the symmetric bilinear form $g(X,Y)$, a Hermitian form $h$ of signature $(n_v,1)$ can be defined from it:
\begin{eqnarray}
h(X,Y) &=& \sp{\ssigma X}{Y}+i \sp{X}{Y}
\, ,
\end{eqnarray}
which can be written in terms of the
projection operators $\Ppm$ as
\begin{eqnarray}
h(X,Y) &=& 2 i \sp{\Pm X}{Y}\nonumber\\
       &=& 2 i \sp{\Pm X}{\Pp Y} \, .
\end{eqnarray}
The three defined structures $\{g, \omega, \ssigma\}$
 form a compatible triple, each structure can be specified
by the two others.



\subsection{$\ssigma$ transformations and Freudenthal duality}

Let us consider 
``$\ssigma$-transformations'' of the type
$$X\to X'=f(\ssigma) X,$$
 where $f$ is an arbitrary 
function. $f(\ssigma)$ can be written with full generality 
as a linear expression 
$$f(\ssigma)\equiv a+b\ssigma$$
 or in ``polar form'' 
$$f(\ssigma)\equiv \lambda \exp(\theta \ssigma)$$
 where $a,b$ or $\lambda,\theta$ are real parameters.
The adjoint is $f(\ssigma)^\dagger=f(\ssigmaad)=f(-\ssigma)$,
$f^\dagger f=a^2+b^2=\lambda^2$. 
Under these transformations $f=f(\ssigma)$ 
the symplectic  and $g$ bilinear products (and then the
Hermitian product $h$
\footnote{Or any other multilinear product built from them}
) become  
scaled:
\begin{eqnarray}
\sp{X'}{Y'}&=&\sp{f X}{f Y}=
\sp{ X}{f^\dagger f Y}=\lambda^2 \sp{ X}{ Y},\\
\sp{\ssigma X'}{Y'}&=&\sp{\ssigma f X}{f Y}=
\sp{ \ssigma X}{f^\dagger f Y}=\lambda^2 \sp{\ssigma X}{ Y}.
\end{eqnarray}
If $\lambda=1$ both products are invariant under the 
Abelian $U(1)$-like group of transformations 
$$U_F(\theta)=\exp \theta\ssigma,$$ 
the  
``$\ssigma$-rotations''. 
Any physical quantity 
(entropy, ADM Mass, scalars at fixed points, intercenter distances,
etc.)
written in terms of these products
(as it will clearly appear in the next sections)
will automatically be scaled under the general 
transformations or invariant under the rotations.

On the other hand
 it can be easily checked that the ``degenerate'' 
Freudenthal duality transformation
\cite{Borsten:2009zy,Ferrara:2011gv,Galli:2012ji,Kallosh:2012yy}. 
is given in our case by  the
the anti-involutive  transformation
\begin{eqnarray}
\tilde{X}&=& \omega \frac{\partial g(X,X)}{\partial X}= 
\ssigma X
\, ,
\end{eqnarray}
with $\tilde{\tilde{X}}=-X$. 

The Freudenthal duality corresponds to a particular 
 $\ssigma$-transformation, a  $\ssigma$-rotation of the
type 
\begin{eqnarray}
f(\ssigma)&=&\exp((\pi/2) \ssigma).
\label{eq7799}
\end{eqnarray}
Invariance of quantities as ADM mass and entropy under 
Freudenthal duality is a special case of a more 
general behavior of the solutions 
under the (Abelian) group of 
general $\ssigma$-transformations.
A general $\ssigma$-transformations can be written 
in terms of Freudenthal duality as
\begin{eqnarray}
X\to  \tilde{X}(a,b)&=& a X+b\tilde{X}, \quad 
\end{eqnarray}
or,
\begin{eqnarray}
 \tilde{X}(\lambda,\theta)&=& \lambda\cos\theta X+\lambda\sin\theta\tilde{X}.
\label{eq7788}
\end{eqnarray}



\subsection{Scalar fields at the fixed points}

Let us turn now to the problem of obtaining the values of the
moduli at the fixed points and at infinity.
 The values of the scalar fields at the fixed points
can  be computed by an explicit expression,
which only involves the matrix $\ssigma_F$.
The fixed values of the $n_v$ complex scalars
$z^\alpha_f(q)$
(at a generic fixed point with charge $q$)
are given, using (\ref{eq226}) and (\ref{eq48}), by
\begin{align}
z^\alpha_f(q)
&=\frac{(\ssigma \I)^\alpha+i\I^\alpha}{(\ssigma\I)^0+i\I^0}
=\frac{\left((\ssigma +i)\I\right)^\alpha}{\left((\ssigma+i\mathds{1})\I\right)^0}
\nonumber\\
&=\frac{(\Pm q)^\alpha}{(\Pm q)^0}
\, .
\label{eq416}
\end{align}
That is, the fixed values of the scalars are given in terms of the projection
of the charges on the ei\-gen\-spa\-ces of the
matrix $\ssigma$.
For quadratic prepotentials, for which $\ssigma$ is a constant,
 this is a complete and explicit solution of the attractor equations.

The values of the $n_v$ complex scalars at spatial infinity,
$|\vecr|\rightarrow\infty$, are given by\footnote{Let us stress that we have used
(\ref{eq226}) and
defined $\I_\infty\equiv \lim_{|\vecr|\to\infty} \I$, but we have not
assumed any particular ansatz for $\I$ up to now.}
\begin{align}
 z^\alpha_\infty
&=\lim_{|\vecr|\rightarrow\infty} \frac{(\Pm \I)^\alpha}
{(\Pm\I)^0}=\frac{(\Pm \I_\infty)^\alpha}{(\Pm \I_\infty)^0}
\, .
\label{eq3434}
\end{align}
According to this formula, the `moduli' $z^\alpha_\infty$
are simple rational functions of the $2 n_v+2$ real constant
components of $\I_\infty$. They are thus independent of the
fixed attractor values \eqref{eq416} (at least for
an $\I$ of interest with only point like singularities,
as for example the ansatz \eqref{eq216}).

We note that  \eqref{eq3434}  is
formally identical to \eqref{eq416}, since they both give the values of the scalars at a fixed point in terms of the charges,
where the roles of $\I_\infty$ and $q$ are exchanged.
It is suggestive then to  write an ``effective
attractor equation'' at infinity, where the center charge  is replaced by the vector $\I_\infty$. That is,
the scalar solutions of the equation
\begin{align}
\I_\infty=\left.\re{2i\bar Z V}\right|_{\infty}
\end{align}
are those ones precisely given  by \eqref{eq3434}.



One can extract some algebraic relations
for the vectors $\I_\infty$ and $q^a$ and the
equations
\eqref{eq416}-\eqref{eq3434}
in specific cases,
for example for solutions with constant scalars.
Let us assume 
$z_f=z_\infty (\not =0) $.
In this case, the equations
\eqref{eq416}-\eqref{eq3434}
 imply the projective equality
($\lambda$ an arbitrary real, non-zero, constant)
\begin{eqnarray}
\Pm \I_\infty&=& \lambda \Pm q\, .
 \end{eqnarray}
which, due to relation \eqref{eq312cc}, implies 
\begin{eqnarray}
 \I_\infty&=& \lambda  q \, .
 \end{eqnarray}
The asymptotic flatness condition
\eqref{eq689a} implies, in addition,
\begin{eqnarray}
\lambda^2&=& \frac{1}{\sp{\ssigma q}{q}}=\frac{1}{2| Z^{f}|^2}
\, .
\end{eqnarray}
The consistency of the last equation is assured by the
positivity of the quadratic form $\sp{\ssigma q}{q}$.
Thus, we can finally arrive to a characterization of $\I_\infty$ in the case of constant scalar solutions
\begin{align}
\I_\infty &= \pm \frac{q}{\sqrt{\sp{\ssigma q}{q}}}\, .
\label{eq118b}
\end{align}
Similar arguments can be stated in the multicenter case.


Let us finish this section with some qualitative remarks.
We have arrived to the expressions \eqref{eq416}-\eqref{eq3434}
which can be written,  in terms of the projective
 complex, vector $\Omega=(X^I,F_I)$, as
\begin{eqnarray}
\Omega_{fix} &=& \Pm q\, , \label{eq445b}\\
\Omega_{\infty} &=& \Pm \I_\infty\, . \label{eq445}
\end{eqnarray}
We could have predicted these expressions a priori:
\footnote{Similar extended  arguments are presented in \cite{Mohaupt:2008gt} and references therein.}
if SUSY  solutions are uniquely determined by the
symplectic real vectors $q_a$, then
the also symplectic but complex vector $\Omega=(X^I,F_I)$
must be related to these vectors in any linear way,
respecting symplectic covariance as well.
Moreover, the symplectic sections $\Omega$ and $V$
lie in the subspace $W^-$, one eigenspace of the
stabilization matrix $\ssigma$.
The only  possibility for such relation
would be the expressions  \eqref{eq445b}-\eqref{eq445},
where the projections of $q$ or $\I_\infty$ on $W^-$ precisely appear.
These expressions, evaluated at the points of maximal symmetry
(the horizon and  infinity), are   equivalent forms of
the  standard horizon attractor equations  and the
generalized attractor equation at infinity presented here.


The scalars at fixed points are invariant not only under
Freudenthal duality or $\ssigma$-rotations 
but also under
general $\ssigma$-transformations of the corresponding
charge vectors. This is clear taking into 
account equations \eqref{eq7778}
and \eqref{eq416}. 
The same conclusion 
applies to the values of the scalars
at infinity (see \eqref{eq3434}) for transformed vectors 
$\I_\infty\to\tilde{\I}_\infty(\lambda,\theta)$.

\section{Complete solutions for quadratic prepotentials}
\label{ssection5}

\subsection{Behavior of the scalar field solutions}

In the previous section we have obtained some general results
without assuming a specific form  for the solutions  $\I$.
In this section we will make use of
the ansatz \eqref{eq216} for theories with quadratic
prepotentials to obtain a full characterization of the
solutions.

Let us insert the ansatz \eqref{eq216} into the general
expression for the complex scalars, \eqref{eq226}.
The values for the time independent $n_v$ complex scalars,
solutions to the field equations, are explicitly given by
\begin{align}
z^\alpha(\vecr)&=
\frac{(\Pm \I)^\alpha}{(\Pm \I)^0}=\frac{(\Pm \I_\infty)^\alpha+\sum_a \frac{(\Pm q_a)^\alpha}{|\vecr-\vecr_a|}}{(\Pm \I_\infty)^0+\sum_a\frac{(\Pm q_a)^0}{|\vecr-\vecr_a|}}
\, .
\label{eq514}
\end{align}
This equation 
is a simple rational expression
for the value of the scalar fields in the whole space.
The fields and their derivatives are regular everywhere,
including the fixed points
(there could be singularities for special charge configurations
which make zero the denominator of \eqref{eq514}).

The expression (\ref{eq514})
 interpolates between the values at the fixed points and at infinity. After some simple manipulations, it can be
written as
\begin{align}
z^\alpha(\vecr)
&=c_\infty^\alpha(\vecr) z^\alpha_\infty +\sum_a c_a^\alpha(\vecr) z^\alpha_{a,f}
\, ,
\label{eq514s}
\end{align}
where
$c_\infty^\alpha(\vecr)$ and $c_a^\alpha(\vecr)$
are spatial dependent complex functions such that
\begin{align}
c_\infty^\alpha(\vecr)+c_a^\alpha(\vecr)
&=
1
\, ,
\nonumber\\
c_\infty^\alpha(\infty)
&=
1
\, ,
\nonumber\\
c_\infty^\alpha(\vecr_a)
&=
0
\, ,
\\
\lim_{\vecr\to\vecr_b}c_a^\alpha(\vecr)&=\delta_{ab}
\, .
\nonumber
\end{align}
For a single center solution, we note that
if $z_\infty^\alpha=z_f^\alpha$ then
the scalar fields are constant in all space.
%

It is straightforward to see that the attractor mechanism
is automatically fulfilled by the ansatz (\ref{eq216}).
The value of $z^\alpha$ at any center $\vecr_a$ is given,
taking the corresponding limit in \eqref{eq514}, by
\begin{align}
z^\alpha(\vecr_a)
&=\frac{(\Pm q_a)^\alpha}{(\Pm q_a)^0}=z^\alpha_f(q_a)
\, ,
\end{align}
where, after the second equality, we have used the
fixed point expression \eqref{eq416},
which is a direct consequence
of the attractor equations.

On the other hand, the solution at the spatial infinity
 recovers spherical symmetry. Again, taking limits, we have
(with $|\vecr|\equiv r$)
\begin{align}
z_\infty^\alpha=z^\alpha( r\to\infty)
&=
\frac{r (\Pm \I_\infty)^\alpha+ (\Pm Q)^\alpha}
{r(\Pm \I_\infty)^0+(\Pm Q)^0}
\nonumber\\
&= (1-c(r))z^\alpha_\infty +c(r)z^\alpha_f(Q)
\, ,
\label{eq514b}
\end{align}
where $z_{f}(Q)$ is the fixed point scalar value which
would correspond, according to the attractor equations,
 to a total charge $Q\equiv \sum_a q_a$.
The asymptotically interpolating function appearing
above is unique for all the scalars
\begin{align}
c(r)&=\frac{1}{1+\frac{r}{r_0}}
\, ,
\end{align}
with the (assumed non-zero) scale parameter
\begin{align}
r_0
&= \frac{(\Pm Q)^0}{(\Pm \I_\infty)^0}
\, .
\label{eq:r0}
\end{align}
They  are such that
\begin{align}
c(0)&=1\, , \quad c(\infty)=0\, .
\end{align}

The \emph{scalar charges} $\Sigma^\alpha$ associated to
the scalar fields can be defined by the
asymptotic series
\begin{align}
z^\alpha( r\to\infty)&=z^\alpha_\infty
+ \frac{\Sigma^\alpha}{r}
+\mathcal{O}\left(\frac{1}{r^2}\right)
\, .
\end{align}
Expanding (\ref{eq514b}), we have
\begin{align}
z^\alpha( r\to\infty)&=z^\alpha_\infty
+ \frac{r_0(  z^\alpha_f(Q)-z^\alpha_\infty)}{r}
+\mathcal{O}\left(\frac{1}{r^2}\right)
\, ,
\end{align}
and thus the scalar charges are given by
\begin{eqnarray}
\Sigma^\alpha&=&r_0\left(  z^\alpha_f(Q)-z^\alpha_\infty\right)
\, .
\label{eq47}
\end{eqnarray}
Hence, the scalar charges are fixed in terms of the charge 
vectors and the asymptotic moduli.
In the special case of a single center solution,
the expression (\ref{eq47}) is in agreement
with the well known fact that the scalar charges vanish
for double extremal black holes.
In the multicenter case, from this formula we infer a
similar result: the scalar charges vanish if
\begin{eqnarray}
z_\infty^\alpha&=&z_{f}^\alpha (Q)
\, .
\label{eq618bb}
\end{eqnarray}
Obviously, this does not mean that the scalar fields
are constant in all space. Therefore, the conditions \eqref{eq618bb} could be considered a convenient generalization of
double extremal solutions in the multicenter case.
Taking into account the considerations of the previous
section, \eqref{eq118b}, a possible vector $\I_\infty$
corresponding to this solution would be of the form
\begin{eqnarray}
\I_\infty&=& \pm\frac{Q}{\sqrt{ \sp{\ssigma Q}{Q}}}
\, ,
\label{eq6179cc}
\end{eqnarray}
and the scalar fields would be parametrized at any point of the space by
\begin{align}
z^\alpha(\vecr)&=c_\infty^\alpha(\vecr) z^\alpha_f(Q)
+\sum_a c_a^\alpha(\vecr) z^\alpha_{ f}(q_a)
\, .
\end{align}
%
%

\subsection{Intercenter distances and $\ssigma$-transformations}
\label{sec42}

The charge interdistances are restricted, from the 
1-form $\omega$ condition of integrability
\cite{Bellorin:2006xr}, we have
(for any charge center $q_b$ )
\begin{eqnarray}
\sp{\I_\infty}{q_b}+\sum_a\frac{\sp{q_a}{q_b}}{r_{ab}} &=&0
\, ,
\label{eq5554}
\end{eqnarray}
where $r_{ab}=| \vecr_a-\vecr_b |$. 
The solutions for this set of equations give the possible values
of the center positions. 

Let us study the effect of $\ssigma$-transformations on the
intercenter distances for transformed $\I_\infty$ and charge
symplectic vectors. The vector $\I_\infty$ is constrained 
by the asymptotic flatness condition to  a unit fixed 
$g$-norm, $1=\sp{\ssigma \I_{\infty}}{\I_\infty}$.
We consider therefore  set of transformations of the type
\begin{eqnarray}
\I_\infty\to \tilde{\I}_\infty(\theta)&=&\exp(\theta \ssigma) \I_\infty,\\
q_a\to \tilde{q}_a(\lambda, \theta)&=&\lambda\exp(\theta \ssigma) 
q_a.
\label{eq5556}
\end{eqnarray}
Under these transformations the equations 
\eqref{eq5554} become
\begin{eqnarray}
\lambda\sp{\I_\infty}{q_b}+\lambda^2\sum_a\frac{\sp{q_a}{q_b}}{\tilde{r}_{ab}} &=&0.
\label{eq5554b}
\end{eqnarray}
Then 
\begin{eqnarray}
r_{ab}\to \tilde{r}_{ab} &=& \lambda r_{ab},
\label{eq5554v}
\end{eqnarray}
the intercenter distances scale (remain invariant) under
general $\ssigma$-transformations ($\ssigma$-rotations or 
Freudenthal dualities) of the charge and $\I_\infty$ vectors.

Let us see the consequence of the integrability equations 
for a double extremal two center configuration. In this 
case, if $\I_\infty=\lambda Q$, we have 
\begin{eqnarray}
0&=&\lambda \sp{Q}{q_2}+\frac{\sp{q_1}{q_2}}{r_{12}}
\nonumber\\
 &=&\lambda \sp{q_1}{q_2}+\frac{\sp{q_1}{q_2}}{r_{12}} 
 \\
 &=&\sp{q_1}{q_2}\left (\lambda+\frac{1}{r_{12}}\right)
 \, .
 \nonumber 
\end{eqnarray}
If we compare this equation with
\eqref{eq6179cc},
we conclude that, if $\sp{q_1}{q_2}\neq 0$,
we have to choose the negative sign there
and the double extremal 
intercenter distance is given by
\begin{eqnarray}
\left.r_{12}^2\right|_{\text{double ext.}} &=&\sp{\ssigma Q}{Q}
\, .
\end{eqnarray}
In the case $\sp{q_1}{q_2}= 0$ the intercenter distance is 
not restristed by the compatibility equation Eq.(\ref{eq5554}).

\subsection{Near horizon and infinity geometry}

Let us now study the  gravitational field.
The   metric has the form given by (\ref{eq211}),
with the asymptotic flatness conditions
$ -g_{rr}=\sp{\R_\infty}{\I_\infty}=1$ and
$\omega(|\vecr|\to\infty)\to 0$.
For point-like sources, as those represented by the
ansatz (\ref{eq216}),
the  compatibility equation (\ref{eq221}) takes the form \cite{Bellorin:2006xr}
\begin{eqnarray}
N  & \equiv& \sum_a\sp{\I_\infty}{q_a}=\sp{\I_\infty}{Q}=0
\, .
\label{eq49}
\end{eqnarray}
 An explicit computation
of the total field strength shows  that (\ref{eq49})
 is equivalent to the requirement of absence of NUT charges:
only  after imposing the condition $N=0$, the overall integral
of the $(F^I,G_I)$ field strengths at infinity
 is equal to $Q=\sum q_a$.
Another consequence of the condition $N=0$, which can
be checked by direct computation from \eqref{eq687bb}, is that
the 1-form $\omega$ takes the same value at any of the horizons of the centers that make up the multicenter black hole. This value is also equal to its value at spacial infinity, which can be taken to be zero.

Let us write  a more  explicit expression for the $g_{rr}$
component at any space point.
We can write, using the `stabilization equation' \eqref{eq888}
and the ansatz \eqref{eq216}, the expression
\begin{eqnarray}
\sp{\R}{\I}&=&
\sp{\ssigma\I_\infty+\sum_a\frac{\ssigma q_a}{|\vecr-\vecr_a|}}{\I_\infty+\sum_b\frac{q_b}{|\vecr-\vecr_b|}}\nonumber\\
&=&1
+\sum_b\frac{1}{|\vecr-\vecr_b|}
\left (\sp{\ssigma\I_\infty}{q_b}+\sp{\ssigma q_b}{\I_\infty}\right )
+
\sum_{a,b}\frac{\sp{\ssigma q_a}{q_b}}{|\vecr-\vecr_a||\vecr-\vecr_b|}
\nonumber\\
&=&1+2\sum_b\frac{\sp{\ssigma\I_\infty}{q_b}}{|\vecr-\vecr_b|}
+\sum_{a,b}\frac{\sp{\ssigma q_a}{q_b}}{|\vecr-\vecr_a||\vecr-\vecr_b|}
\, ,
\label{eq222}
\end{eqnarray}
where we have  used  the property
$\ssigma^\dagger=-\ssigma$ and the
asymptotic flatness
condition $\sp{\ssigma \I_\infty}{\I_\infty}=1$.
We  introduce now the quantities
\begin{align}
 M_a        &\equiv \sp{\ssigma \I_\infty}{q_a}\, , \label{eq432}\\
 A_{ab} &\equiv \sp{\ssigma q_a}{q_b}\, , \label{eq433}
\end{align}
where $A_{ab}$ is symmetric in its indices
due to the property \eqref{eq1123b}.

With these definitions, we finally write the
expression for the metric element as
\begin{align}
-g_{rr}&=\sp{\R}{\I}
\nonumber\\
&=
1+2\sum_a\frac{M_a}{|\vecr-\vecr_a|}
+\sum_{a,b}\frac{  A_{ab}}{|\vecr-\vecr_a||\vecr-\vecr_b|}
\, .
\label{eq415}
\end{align}
If the metric element (\ref{eq415}) describes a
black hole,  then the right part should be kept
 always positive  and  finite for any finite $|\vecr|$.\footnote{Consider, for example, that $-g_{rr}\sim e^{-\K}>0$.} 
A sufficient condition for  its positivity is, for example, 
that  the mass-like $M_a$ and area-like $A_{ab}$ parameters 
are all positive. 


\subsubsection*{Behavior at fixed points and at infinity}

We will define new quantities from the behavior of the metric
at infinity: the mass
$M_{ADM}$ and $\atot$.
At spatial infinity $|\vecr|\to \infty$, $1/| \vecr-\vecr_a|\to 1/r$, the metric element \eqref{eq415}
becomes  spherically symmetric:
\begin{align}
-g_{rr}
&=1+\frac{2\sum_a M_a}{r}+ \frac{ \sum_{ab}A_{ab}}{r^2}+\mathcal{O}\left(\frac{1}{r^3}\right)
\nonumber\\
&\equiv
1+\frac{2M_{ADM}}{r}+ \frac{ \atot}{r^2}+\mathcal{O}\left(\frac{1}{r^3}\right)
\, .
\end{align}
The second equation defines $M_{ADM}$ and \atot.
Comparing both expressions and using
\eqref{eq432}, \eqref{eq433}, we have
\begin{align}
M_{ADM} &= \sum_a M_a=\sp{\ssigma I_\infty}{Q}\, ,
\label{eq529}\\
\atot &= \sum_{ab} A_{ab} =\sp{\ssigma Q}{Q}\, .
\label{eq530}
\end{align}
The expression for the central charge at
infinity, (\ref{eq218b}),
becomes then
\begin{eqnarray}
| Z_\infty |^2 &=& \mid \sp{\Pp \I_\infty}{Q} \mid^2=\mid N+ i M\mid^2\\
              &=& M_{ADM}^2+N^2
\label{eq531b}
\end{eqnarray}
where $N$ is defined by
(\ref{eq49}).
The compatibility
condition $N=0$  is equivalent to the saturation
of a BPS condition
\begin{align}
| Z_\infty |^2 &= M_{ADM}^2 =| \sp{\ssigma \I_\infty}{Q}|^2
\, .
\label{eq531bc}
\end{align}

Unlike $\atot$, the $M_{ADM}$ quantity 
depends on the values of the scalars at infinity
through the implicit dependence of $\I_\infty$ on them.
These can take arbitrary values or, at least, can be
chosen in a continuous range.
In the single center case, for any given charge vector, one
can obtain a certain particular solution by setting the
scalar fields to constant
values ($z_{f}^\alpha=z^\alpha_\infty$), giving this the minimal possible $M_{ADM}$ mass
\cite{Gibbons:1996af}.
For multicenter solutions and generic non-trivial
charge vectors, it is not possible to have constant scalar
fields. Nevertheless, we can still proceed to 
the extremization of $M_{ADM}(z_\infty^\alpha)$,
with respect to the scalar fields at infinity for a given
configuration. 


Let us suppose a configuration with null scalar charges. 
In this case $\I_\infty=\pm \lambda Q$, 
$\lambda=1/\sqrt{\sp{\ssigma Q}{Q}}$. We have 
$M_{ADM}=\pm/\lambda$, the positivity of $M_{ADM}$ obliges us to 
choose the positive sign. For a two center case
 this in turn implies $\sp{q_1}{q_2}=0$ 
 and $r_{12}$ unrestricted (see section (\ref{sec42})).  
Such $\I_\infty$ trivially satisfies the absence of 
NUT charge ($N=0$) condition,
and  for it $z_\infty^\alpha=z_{f}(Q)$. This configuration
can be considered a multicenter generalization of the double extremal solutions.

Let us proceed now with the study of the geometry near the centers.
For $\vecr\to \vecr_a$ the metric element
given by  (\ref{eq415}) becomes spherically symmetric.
Moreover, it can be shown that,
by fixing additive integration constants,
we can take $\omega_a=\omega(\vecr\to\vecr_a)=0$
at the same time that $\omega_\infty=\omega(\vecr\to\infty)=0$.
As a consequence, the metric at any of the horizon components
with charge $q_a$
approaches an $AdS_2\times S^2$ metric of the form
\begin{align}
ds^2
&= \frac{r^2}{\sp{\ssigma q_a}{q_a}} dt^2- \frac{\sp{\ssigma q_a}{q_a}}{r^2} d{\vecr}^2
\, .
\label{eq425}
\end{align}
This is a Robinson-Bertotti-like metric. The Robertson-Bertotti-like mass parameter $M_{RB}$ is given by
\begin{align}
M_{RB,a}^2 &=\sp{\ssigma q_a}{q_a},
\end{align}
this is a charge extremal condition impliying
the positivity of the charge products: 
$\sp{\ssigma q_a}{q_a}> 0$.
\footnote{The positivity of these quantities implies 
diverse restrictions as the quadratic form 
$g(X,Y)=\sp{\ssigma X}{Y}$ is undefinite with a signature 
including an even number of negative signs.}

Then, the near horizon geometry is 
completely determined in terms of the individual
horizon areas $\ah_{,a}=\sp{\ssigma q_a}{q_a}$.
The horizon area $\ah$ is the sum
of the areas of its disconnected parts
\begin{align}
\ah
=\sum_a \ah_{,a} 
= \sum_a \sp{\ssigma q_a}{q_a}
=2 \sum_a | Z_{f,a}|^2
\, .
\end{align}
This expression can be compared with the
area corresponding to a single center black hole with
the same total charge
$Q=\sum_a q_a$, which is given by
$\ah(q=Q)=\sp{\ssigma Q}{Q}$.


The relation between the asymptotic ``area'' $\atot$ and
the multicenter horizon area, or horizon entropy $\ah$, is 
simply
\begin{eqnarray}
\atot &=&\sp{\ssigma Q}{Q}= \sum_{a,b} \sp{\ssigma q_a}{q_b}\nonumber\\
&=& \ah+2\sum_{a< b} \sp{\ssigma q_a}{q_b}\, .
\label{eq429}
\end{eqnarray}
For one center solution we always have $\atot=\ah$.
For example, in the case of two centers  with
charges $q_1$, $q_2$, the difference is\footnote{In ref. \cite{Ferrara:2010cw} it has been shown  that, for quadratic prepotentials, the single center BPS extremal black hole area with charge $Q=q_1+q_2$ is always larger than the corresponding two-center area 
\begin{eqnarray*}
\ah(Q=q_1+q_2)&\geq&\ah_{,1}+\ah_{,2}
\, .
\end{eqnarray*}
Or equivalently, taking into account that $\atot$ is also the area of the corresponding single center black hole with the same total 
charge $\atot_{q1,q2}=S_h(Q=q_1+q_2)$, we have
\begin{eqnarray*}
\atot-\ah&=& 2 \sp{\ssigma q_1}{q_2}\ge 0
\, .
\end{eqnarray*}}
\begin{eqnarray}
\atot-\ah&=& 2 \sp{\ssigma q_1}{q_2}>0\, .
\label{eq430}
\end{eqnarray}

Let us finally remark that 
under  $\ssigma$-transformations 
$\I_\infty\to \tilde{\I}_\infty(\theta)$, 
$q_a\to \tilde{q}_a(\lambda, \theta)$
the $ADM$ mass and the horizon areas scale as
\begin{eqnarray}
\tilde{M}_{ADM} &=&\lambda M_{ADM}\, ,\\
\tilde{\ah} &=&\lambda^2 \ah\, .
\end{eqnarray}
Under the same transformations, the 
scalars at the fixed points and at infinity remain invariant whereas
the
intercenter distances
$r_{ab}$ scale as \eqref{eq5554v}.

\section{Freudenthal duals and charge vector expansions}
\label{ssection6}




It is well known the  utility of the use of the section 
$\V$, its derivatives $D_\alpha \V$ and their complex conjugates as 
a basis for the symplectic space. 
Any real symplectic vector $X$ can be expanded as
\begin{align*}
X =2\im{Z(X)\bar{V}+g^{\alpha\bar\beta}D_\alpha Z(X)\bar{D}_{\bar \beta} \bar{V}}
\end{align*}
with $Z(X)=\sp{V}{X}$.
The existence and properties of such
 expansions are based on the symplectic properties of 
$\V$ and its derivatives as well as on the existence of a anti-involution $\ssigma(N)$ for which 
$\ssigma(N)\V= i\V$ and $\ssigma(N)D_\alpha\V= iD_\alpha\V$.\footnote{See for example Section 2.2.2 in \cite{thesiskatmadas} and references therein.}

We will define here alternative expansions using the 
properties of the matrix $\ssigma\equiv \ssigma(F)$.
As we have seen before, the projectors $\Ppm$ split the $(2n_v+2)$-dimensional space $W$ into two $(n_v+1)$-dimensional eigenspaces
\begin{align*}
 W=W^+\oplus W^- 
 \, ,
\end{align*}
in which the eigenvectors of $\ssigma$
(therefore eigenvectors of general $\ssigma$-transformations)
 with eigenvalues $\pm i$, respectively, lie (\emph{cf.} Section 3).

Given a set of generic real charge vectors $(q_1,\ldots,q_{n})$, the sets  $(\Pp q_a)$, respectively $(\Pm q_a),$ possibly completed with additional suitable vectors, can be considered a
basis for the eigenspaces $W^+$, respectively $W^-$.
Let us consider the $W$  subspace  $B(q_n)$ generated 
by eigenvectors of the matrix $\ssigma$ associated 
to  center charges, directly of the complex form
\begin{align}
B(q_n)\equiv \text{Span}(\Ppm q_1,\ldots,\Ppm q_n)\, ,
\end{align}
or, equivalently, in the real basis
formed by charge vectors and their 
Freudenthal duals $\tilde{q}_i=\ssigma q_i$
\begin{align}
B(q_n)\equiv \text{Span}(q_1,\ldots,q_n,\ssigma q_1,\ldots,\ssigma q_n)
\, .
\end{align}
In particular, we can consider the subspace $B(q_{na})$
generated by the $n_a$ pairs ($q_a$, $\ssigma q_a$)
of center charges, whose dimension is, in general,
$\dim B(q_{na})\leq 2 n_a$.
The dimension  of the orthogonal complement to this
space, $B(q_{na})^\perp$,
\emph{i.e.} those vectors $s$ such
that $\sp{q}{s}=\sp{\ssigma q}{s}=0$
is, generically, $\dim B(q_{na})^\perp= 2 (n_v-n_a)+2$.\footnote{Or equivalently, $B(q_{na})^\perp$ is defined as the
set of vectors $s$ such that
$h(s,q)=0$ for all $q\in (q_{na})$, where $h$ is the Hermitian
inner product defined in Section \protect\ref{ssection4}.}
This dimension  is zero for one scalar, one
center black holes ($n_v=0,n_a=1$).
The set of vectors ($q_a$, $\ssigma q_a$)
may form themselves a (maybe overcomplete) basis for the
$(2 n_v+2)$ symplectic space. Otherwise, they
can be extended with as many other vectors ($s_i$)
as necessary to complete such basis.
Any real symplectic vector of interest  (\emph{e.g.} $\I_\infty$) 
can be conveniently expanded as
\begin{eqnarray}
X &=& 2 \Re \alpha^a \Pp q_a+2\Re \gamma^i \Pp s_i
\, ,
\end{eqnarray}
where $\alpha^a,\gamma^i$ are complex parameters 
or, equivalently, as
\begin{eqnarray}
X &=&  \alpha^a  q_a+\tilde{\alpha}^a\ssigma q_a+ \gamma^i  s_i+
\tilde{\gamma}^i \ssigma s_i
\, ,
\end{eqnarray}
where $\alpha^a,\tilde{\alpha}^a,\gamma^i,\tilde{\gamma}^i$ 
are in this case real parameters.\footnote{Naturally, other bases are possible or convenient, as for example bases including linear combinations of the charge vectors, the total charge vector $Q$, $\I_\infty$, etc.}
Let us note that under this same expansion 
the dual vector $\tilde{X}=\ssigma X$ 
has respectively complex components 
$(-i\alpha,\dots)$ or real 
ones 
$(-\tilde{\alpha}^a,\alpha^a,\dots  )$.

We can use  expansions of different quantities
in such  a basis formed by charge and extra vectors,
to get different results.
In a simple illustrative case, by decomposition of 
the $\I_\infty$ vector, we will study different properties. 
In particular, we will see how the extremality of the 
solutions imposes strong conditions on such extra vectors.
%


\subsection{Decomposition of $\I_\infty$ and double extremality}

We will decompose now the vector $\I$ into a basis of
charge and extra vectors. For the sake of simplicity we
will discuss here the case of
a single center solution and one complex scalar.
The dimension of the symplectic space is  $2 n_v+2=4$.
We will see, in
particular, how the extremality of the solutions imposes
strong conditions on such extra vectors.
In addition, we will show, using this decomposition,
the double extremality of the black hole solutions
 for quadratic prepotentials.

Let us decompose the vector $\I_\infty$ in the following way
(with  $\sp{\ssigma q}{q}\neq0$):
\begin{eqnarray}
\I_\infty
&=&\alpha q+\beta\ssigma q+\gamma s +\epsilon \ssigma s\, ,
\label{eq51}
\end{eqnarray}
where $\alpha,\beta,\gamma,\epsilon \in\mathbb{R}$
and $s \in B(q_{na},\ssigma q_{na})^\perp$ is arbitrary. 
The bilinear form $g(X,Y)=\sp{X}{Y}$ 
is indefinite, it has
a signature with an even number of minus signs. 
A physical requirement is that  
$\atot=\sp{\ssigma q}{q}>0$
that implies that, if we choose
\begin{align}
 \sp{s }{q}        &=\sp{s}{\ssigma q}=0\,
\end{align}
then \footnote{By a simple application of the Silverster inertia 
theorem.} we are obliged to choose
\begin{align}
 \sp{\ssigma s }{s }&= -1.
\end{align}
This vector $s$ can  be always determined by a modified
 Gram-Schmidt procedure for a given pair of
vectors $(q,\ssigma q)$.
By projecting the relation (\ref{eq51}) over any of the
individual vectors $(q,\ssigma q)$,
we get
\begin{eqnarray}
\sp{\I_\infty }{q} &=& \beta \sp{\ssigma q}{q}\, ,\\
\sp{\I_\infty }{\ssigma q} &=& -\alpha \sp{\ssigma q}{q}\, .
\end{eqnarray}
Using  \eqref{eq49}, \eqref{eq529} and \eqref{eq530},
we can rewrite these last two expressions respectively as
\begin{eqnarray}
N&=& \beta \atot\, ,\label{eq56}\\
M_{ADM}&=& -\alpha \atot \, ,\label{eq57}
\end{eqnarray}
from which we read the coefficients $\alpha$, $\beta$ in terms of some other, more physical, parameters.
The condition $N=0$ implies that $\beta=0$, hence the
$\I_\infty$ vector does not contain any component in the
``$\ssigma q$'' direction.

Let us consider now the asymptotic flatness condition
and apply the ansatz (\ref{eq51}) for $\I_\infty$,
but without imposing at this moment the $N=0$ condition.
If we define $\Delta^2\equiv \gamma^2+\epsilon^2$ and make 
use of \eqref{eq531b} and the values for $\alpha$, $\beta$, we have
\begin{align}
1&=\sp{\ssigma\I_\infty}{\I_\infty}\nonumber\\
&=\left(  \alpha^2+\beta^2\right)\sp{\ssigma q}{q}+
\left(  \gamma^2+\epsilon^2\right)\sp{\ssigma s}{s}\nonumber\\
&=
\frac{M_{ADM}^2+N^2}{\atot^2}\sp{\ssigma q}{q}-\Delta^2\,  ,
\end{align}
or equivalently,
\begin{align}
| Z_\infty |^2=M_{ADM}^2+N^2
&=\sp{\ssigma q}{q}(1+\Delta^2)
\, .
\end{align}
The BPS condition
$| Z_\infty |=M_{ADM}=\sp{\ssigma q}{q}$ is only fulfilled
if $N=0$ (in agreement with \eqref{eq531bc})
and the additional condition $\Delta=0$. 

The vanishing of these quantities can be directly
seen  by
imposing extremality in the metric elements,
 by requesting extremal RN black hole type metric or,
$-g_{rr}\sim f^2$ with $f$ an spatially harmonic function.
The metric component $g_{rr}$ is
\begin{align}
-g_{rr}
&=  
1+\frac{2M_{ADM}}{r}+\frac{\sp{\ssigma q}{q}}{r^2}
\nonumber\\
&=
1+\frac{2M_{ADM}}{r}+\frac{(M_{ADM}^2+N^2)/(1+\Delta^2)}{r^2}
\nonumber\\
&=
\left (   1+\frac{M_{ADM}}{r}\right )^2+\frac{1}{r^2}
\frac{1}{1+\Delta^2}
\left(M_{ADM}^2\Delta^2+ N^2 \right)
\, .
\end{align}
The metric element is of the form
$-g_{rr}\sim f^2$ with $f$ an spatially harmonic function
if and only if the second term of the previous expression is
zero, that is, if and only if
\begin{align}
M_{ADM}^2\Delta^2+ N^2
&=0
\, .
\end{align}
Thus, the conditions
$N=0$ and $\Delta=0$
(which is equivalent to $\gamma=\epsilon=0$ in \eqref{eq51})
are necessary conditions to
recover an extremal RN black hole type metric.
In this case, the central charge at infinity is
\begin{align}
| Z_\infty |^2 &= M_{ADM}^2 =\sp{\ssigma q}{q}\, .
\label{eq616bc}
\end{align}

We see that the vanishing of the non-extremality
parameter $\Delta$ is equivalent to
require that   $\I_\infty$ is fully contained in the
subspace $\text{Span}(q,\ssigma q)$, whereas the condition
$N=0$ further restricts it to be proportional to  the
vector charge $\I_\infty=q/M_{ADM}$.
In this case, after imposing the conditions $N=\Delta=0$,
we can finally write
\begin{align}
\I  =& \frac{q}{M_{ADM}}\left (1+ \frac{M_{ADM}}{r}\right)
\, .
\label{eq567}
\end{align}
As a consequence of having $\I_\infty=q/M_{ADM}$ the scalar fields $z^\alpha$ are constant everywhere
and equal to their values at the fixed point (see \eqref{eq3434} and the discussion in 
Section \ref{ssection5}). It might be interesting to remark that in this expression the ``unphysical'' 
vector $\I$ appears written in terms of the physical quantities $q$ and $M_{ADM}$ which can be chosen by 
hand from the beginning.

\section{Summary and concluding remarks}
\label{ssection7}

We have presented a systematic study of general, stationary, multicenter black hole solutions in $N=2$ $D=4$  Einstein-Maxwell  supergravity theories minimally coupled to  scalars, \emph{i.e.}
theories with quadratic prepotentials.
We have assumed a generic multicenter ansatz \eqref{eq216}, which depends on $q_a$ and $\I_\infty$.

This analysis is heavily  based on the use of the algebraic 
properties of the  anti-involutive matrix $\ssigma$, the 
constant matrix of second derivatives of the prepotential 
of the theory and of the, defined in this work, symplectic 
adjoint $\ssigmaad$.
They are  ``unitary'', $\ssigma \ssigmaad=\mathds{1}$, with respect to the symplectic product. 
The matrix $\ssigma$ defines a complex structure
on the $(2n_v+2)$-dimensional space symplectic space.
%

By defining suitable projector operators $\Ppm$, the symplectic 
$(2n_v+2)$-dimensional space is decomposed into 
eigenspaces of  $\ssigma$, $W=W^+\oplus W^-$. 
We show that any symplectic section, whose real and imaginary parts are related  $\re{X}=\ssigma\im{X}$, lies in the 
subspace $W^-$.
With the help of these projector operators, we write a 
purely algebraic expression for the attractor equations, 
which equalizes the symplectic section $\V\in W^-$ 
to the projection of the corresponding 
charge vector on that subspace, 
$\Pm q^a\in W^-$ \eqref{eq48}-\eqref{eq311zz}. 
The modulus of the central charge function is given in terms 
of the norm of a charge vector, which is written in terms of the inner product $g$, \eqref{eq412}.

We obtain expressions for the scalar fields evaluated at the 
fixed points \eqref{eq416} and at infinity \eqref{eq3434}. 
They are given, in a similar way, in terms of the projections 
of the center charges vectors $q_a$ and $\I_\infty$ on $W^-$, respectively.
The values of the $n_v$ complex scalars at spatial infinity
 are given by \eqref{eq3434}
\begin{align}
 z^\alpha_\infty
&=\lim_{|\vecr|\rightarrow\infty} \frac{(\Pm \I)^\alpha}
{(\Pm\I)^0}=\frac{(\Pm \I_\infty)^\alpha}{(\Pm \I_\infty)^0}
\, .
\end{align}
This is an explicit formula where the moduli $z^\alpha_\infty$
are simple rational functions of the $2 n_v+2$ real constant
components of $\I_\infty$.

We  write expressions for the scalar field solutions 
 at any space point \eqref{eq514s} in terms of $q_a$ and $\I_\infty$ \eqref{eq:r0}. They are interpolating expressions between
the fixed point values and moduli values at infinity.
In particular, the formalism allow us to easily 
  study a  configuration analogous to the 
\emph{double extremal} case in a multicenter scenario:
 configurations such that $z_\infty^\alpha=z_{f}^\alpha (Q)$,
with  $Q$ the total charge. 
The vanishing of the scalar charges  is shown to be equivalent 
to this condition. This is in close analogy with the single center case, in which the vanishing of the
scalar charges is a necessary and sufficient condition for the double extremality of the black hole \cite{Gibbons:1996af}.

We have written the metric element $-g_{rr}$ in terms of 
area-like $A_{ab}$ and mass-like quantities
 $M_a$ \eqref{eq415} involving the bilinear product $g$. 
The study of the near horizon and infinity geometry of the solution lead us to the consideration of  the area-like quantities
$A_{ab}=\sp{\ssigma q_a}{q_b}$ and 
$\atot=\sum_{ab} A_{ab}=\sp{\ssigma Q}{Q}$,
in addition to the horizon areas
$\ah_{,a}=\sp{\ssigma q_a}{q_a}$.

In Section \ref{ssection6} we have proposed a decomposition
of the $(2 n_v+2)$-dimensional symplectic 
 vector space in a basis of  
eigenvectors of the matrix $\ssigma$.  This
 set of vectors are of the form
$(\Ppm q_a)$, 
or, alternatively, $(q_a,\ssigma q_a)$,   
with $\Ppm$  projectors over the eigenspaces of $\ssigma$
and $q_a$ the center vector charges. 
Any real symplectic vector of interest  (\emph{e.g.} $\I_\infty$) can be conveniently expanded as
($\alpha^a,\gamma^i$ are complex parameters )
\begin{eqnarray}
X &=& 2 \Re \alpha^a \Pp q_a+2\Re \gamma^i \Pp s_i
\, ,
\end{eqnarray}
or  as
( real $\alpha^a,\tilde{\alpha}^a,\gamma^i,\tilde{\gamma}^i$) 
$X =  \alpha^a  q_a+\tilde{\alpha}^a\ssigma q_a+ \gamma^i  s_i+
\tilde{\gamma}^i \ssigma s_i.$
Some simple properties
of the solutions are studied using this decomposition.
The decomposition can be seen an alternative to 
the well known expansions in terms of the section 
$\V$, its derivatives $D_\alpha \V$ and their complex conjugates as  a basis for the symplectic space. A formalism which allows
that any real symplectic vector $X$ can be expanded as
$X =2\im{Z(X)\bar{V}+g^{\alpha\bar\beta}D_\alpha Z(X)\bar{D}_{\bar \beta} \bar{V}}$

The anti-involution matrix $\mathcal{S}$ can 
be understood as a  Freudenthal duality $\tilde{q}=\ssigma q$
\cite{Borsten:2009zy,Ferrara:2011gv}. 
Under this transformation of the charges the horizon area,
 ADM mass  and other properties of the solutions remain invariant.
We have shown, for the quadratic prepotential theories studied here, that this duality can be generalized to an Abelian 
 group of transformations (``Freudenthal transformations'') 
of the form 
$$x\to \lambda\exp(\theta \ssigma) x= a x+b\tilde{x}.$$
Under this set of transformations applied to 
the charge vectors and $\I_\infty$, 
 the horizon area, ADM mass and intercenter 
distances scale up, respectively, as 
\begin{equation}
\ah\to\lambda^2\ah,\quad M_{ADM}\to\lambda M_{ADM},\quad 
r_{ab}\to\lambda r_{ab} ,
\end{equation}
leaving invariant the values of the scalars at the fixed points and at infinity.
In the special case $\lambda=1$, ``$\ssigma$-rotations'', 
the transformations leave invariant the solution. 
The standard Freudenthal duality can be written as
the particular rotation
 $$\tilde x= \exp(\pi/2 \ssigma) x .$$

It is immediate to ask the question whether such 
transformations can be generalized to $4d$ theories with 
general prepotentials, not associated to ``degenerate'' U-duality groups, including stringy black holes.
We can see that this is indeed the case using a simple 
argument as follows (a more detailed investigation is
presented in \cite{torrentejordan}). 
The U-duality quartic invariant defined 
(\cite{Borsten:2009zy}, using here a slightly adapted notation
) as
$$ 2\Delta_4(x)\equiv \sp{T(x)}{x}$$
 can be written also, 
using the definition of Freudenthal duality, as
$$ \Delta_4(x)= \frac{1}{4}\sp{\tilde{x}}{x}^2.$$ 
Let us note then that, for a general transformation
this quantity scale as
\begin{eqnarray}
 2\Delta_4( a x+b \tilde{x})^{1/2} &=&
\sp{\widetilde{ a x+b \tilde{x}}}{ a x+b \tilde{x}}=
\sp{a \tilde{x}-b x}{ a x+b \tilde{x}}\\
&=&(a^2+b^2)\sp{ \tilde{x}}{x}\\
&=&2(a^2+b^2)\Delta_4( x)^{1/2}. 
\label{eq9977ww}
\end{eqnarray}
For $a^2+b^2=1$, a $\ssigma$-rotation, the 
quantity $\Delta_4$ for any $U$-duality group, and 
then the lowest order entropy of any 
extremal stringy black hole, is invariant under these 
transformations. 

Moreover the invariance of $\Delta_4$ is shown  \cite{torrentejordan}
to be equivalent to the conditions 
\begin{eqnarray}
\Delta_4(x,x,\tilde{x},\tilde{x})&=&\frac{1}{3}\Delta_4(x),
\label{eq2211zz1}\\ 
\Delta_4(x,\tilde{x},\tilde{x},\tilde{x})&=&
\Delta_4(x,x,x,\tilde{x})=0.
\label{eq2211zz2}
\end{eqnarray}
For the special case of 
$D=4$ theories with  U-duality groups of ``degenerate type $E_7
$''
such conditions (\ref{eq2211zz1}-\ref{eq2211zz2})can be 
easily checked by an explicit 
computation.

\vspace{0.3cm}
\section*{Acknowledgements}
We acknowledge T. Ort\'{\i}n for many useful comments and suggestions
along the preparation of this work.
This work has been supported in part by the Ministerio de 
Educaci\'on y Ciencia grants FIS2011-3454, FPA2008-2356B and
the Universidad de Murcia project E024-018.
The work of JJ.F-M has been supported by a FPI-predoc 
contract FPI-2009-2132.
We thank the hospitality of the CERN TH Division and of the
IFT-CSIC (Madrid) where part of the research has been conducted.

\end{document}